\documentclass[sigconf]{acmart}

\copyrightyear{2024}
\acmYear{2024}
\setcopyright{rightsretained}
\acmConference[MSR '24]{21st International Conference on Mining Software Repositories}{April 15--16, 2024}{Lisbon, Portugal}
\acmBooktitle{21st International Conference on Mining Software Repositories (MSR '24), April 15--16, 2024, Lisbon, Portugal}\acmDOI{10.1145/3643991.3644884}
\acmISBN{979-8-4007-0587-8/24/04}

\usepackage[T1]{fontenc}
\usepackage{graphicx} %
\usepackage{xspace}
\usepackage{paralist}
\usepackage{hyperref}
\usepackage{tablefootnote}
\usepackage{graphicx}
\usepackage{booktabs}
\usepackage{color}
\usepackage{tabularray}
\usepackage{natbib}
\usepackage{balance}

\newcommand{\TODO}[1]{\textcolor{red}{#1}\GenericWarning{}{LaTeX Warning: TODO: #1}}\newcommand\todo\TODO

\newcommand{\name}{\textsc{GitBug-Java}\xspace}

\begin{document}

\title{\name: A Reproducible Benchmark of Recent Java Bugs}

\author{André Silva$^*$}
\affiliation{
    \institution{KTH Royal Institute of Technology}
    \city{Stockholm}
    \country{Sweden}
}
\email{andreans@kth.se}

\author{Nuno Saavedra$^*$}
\affiliation{
    \institution{INESC-ID/IST, University of Lisbon}
    \city{Lisbon}
    \country{Portugal}
}
\email{nuno.saavedra@tecnico.ulisboa.pt}

\author{Martin Monperrus}
\affiliation{
    \institution{KTH Royal Institute of Technology}
    \city{Stockholm}
    \country{Sweden}
}
\email{monperrus@kth.se}

\begin{abstract}
Bug-fix benchmarks are essential for evaluating methodologies in automatic program repair (APR) and fault localization (FL). 
However, existing benchmarks, exemplified by Defects4J, need to evolve to incorporate recent bug-fixes aligned with contemporary development practices. 
Moreover, reproducibility, a key scientific principle, has been lacking in bug-fix benchmarks.
To address these gaps, we present \name, a reproducible benchmark of recent Java bugs.
\name features 199 bugs extracted from the 2023 commit history of 55 notable open-source repositories.
The methodology for building \name ensures the preservation of bug-fixes in fully-reproducible environments.
We publish \name at \url{https://github.com/gitbugactions/gitbug-java}.
\end{abstract}

\keywords{Software Bugs, Bug Benchmark, Reproducibility, Bug Database, Java Benchmark, Software Testing, Program Analysis}

\maketitle

\def\thefootnote{*}\footnotetext{These authors contributed equally to this work.}\def\thefootnote{\arabic{footnote}}

\section{Introduction}

Bug-fix benchmarks are integral to advancing the field of software engineering, providing crucial resources for soundly evaluating research in sub-fields such as automatic program repair (APR) and fault localization (FL) \cite{sim2003using, wright2010validity, just2014defects4j}.
These benchmarks capture the software modifications made to rectify existing defects, aligning program behavior with the intended specifications.
For example, Defects4J \cite{just2014defects4j} is a widely adopted benchmark that has significantly contributed to software engineering research over the past decade. Yet, most bugs from Defects4J are between 10 and 15 years old.
In this paper, we address two problems related to current benchmarks:

\emph{Problem 1: Bugs are not recent.} It is of utmost importance that benchmarks incorporate recent bug-fixes.
This ensures relevance of evaluations with respect to current software stacks and development practices.
It also mitigates the risk of data leakage, a concern when evaluating large language models (LLMs) using data seen during training \cite{jacovi2023stop, zhang2023critical, lee2023github}.
In the context of bug-fix benchmarks, data leakage refers to the fact that some machine learning code models have seen the code from the benchmark at training time. Data leakage, or benchmark leakage ruins scientific validity.

\emph{Problem 2: Bugs are not reproducible.} Next, the reproducibility of benchmarks is crucial for the replication of studies as well as for systematic comparative evaluations in subsequent research.
For bug-fix benchmarks, reproducibility means the ability of compiling and running the code for each bug, even years and decades after their initial appearance.
Despite the fundamental role of reproducibility in the scientific method, bug-fix benchmarks have struggled to maintain it over time.
Indeed, Zhu et al.  \cite{zhu2023reproducibility} showed that reproducibility in bug-fix benchmarks ranges between 26.6\% and 96.9\%, with none achieving full reproducibility.
In this paper, we aim at addressing the two challenges of recency and reproducibility in the context of Java.
\emph{We contribute with a high-quality, reproducible bug-fix benchmark only composed of recent Java bugs.}

We present \name, a reproducible benchmark of recent Java bugs.
\name is built using the framework GitBug-Actions \cite{saavedra2023gitbug} to identify and select high quality bug-fixes.
We use the development environment and build pipelines defined by developers in their GitHub Actions to locally execute commits from open-source repositories.
We only select commits from 2023, which is after the cut-off date of the training data from all notable LLMs, including OpenAI models.
We package these bug-fixes in fully-reproducible environments by storing all necessary files, including all software dependencies.

\name contains a total of 199 bugs collected from 55 different relevant open-source repositories.
All bugs in \name are from the year of writing, 2023: this addresses the need for recent bugs.
All bugs in \name ship with reproduction environments that can be executed offline: this guarantees reproducibility.

In summary, our contributions are:
\begin{itemize}
    \item A sound methodology for benchmark construction that ensures 1) recency of collected bug-fixes 2) long-term execution reproducibility of the bugs and their fixes.

    \item \name, a reproducible benchmark of recent Java bugs containing 199 bugs from 55 repositories: 1) \name facilitates future research on program repair, fault localization, and related fields; 2) \name addresses the crucial problem of benchmark leakage in LLM-based research.

    \item A repository on GitHub that 1) is made publicly available at \url{https://github.com/gitbugactions/gitbug-java}; 2) comes with proper documentation and a visualization companion website.

\end{itemize}

\newpage
\section{Building \name}

\begin{figure}[t]
    \centering
    \includegraphics[width=0.7\linewidth]{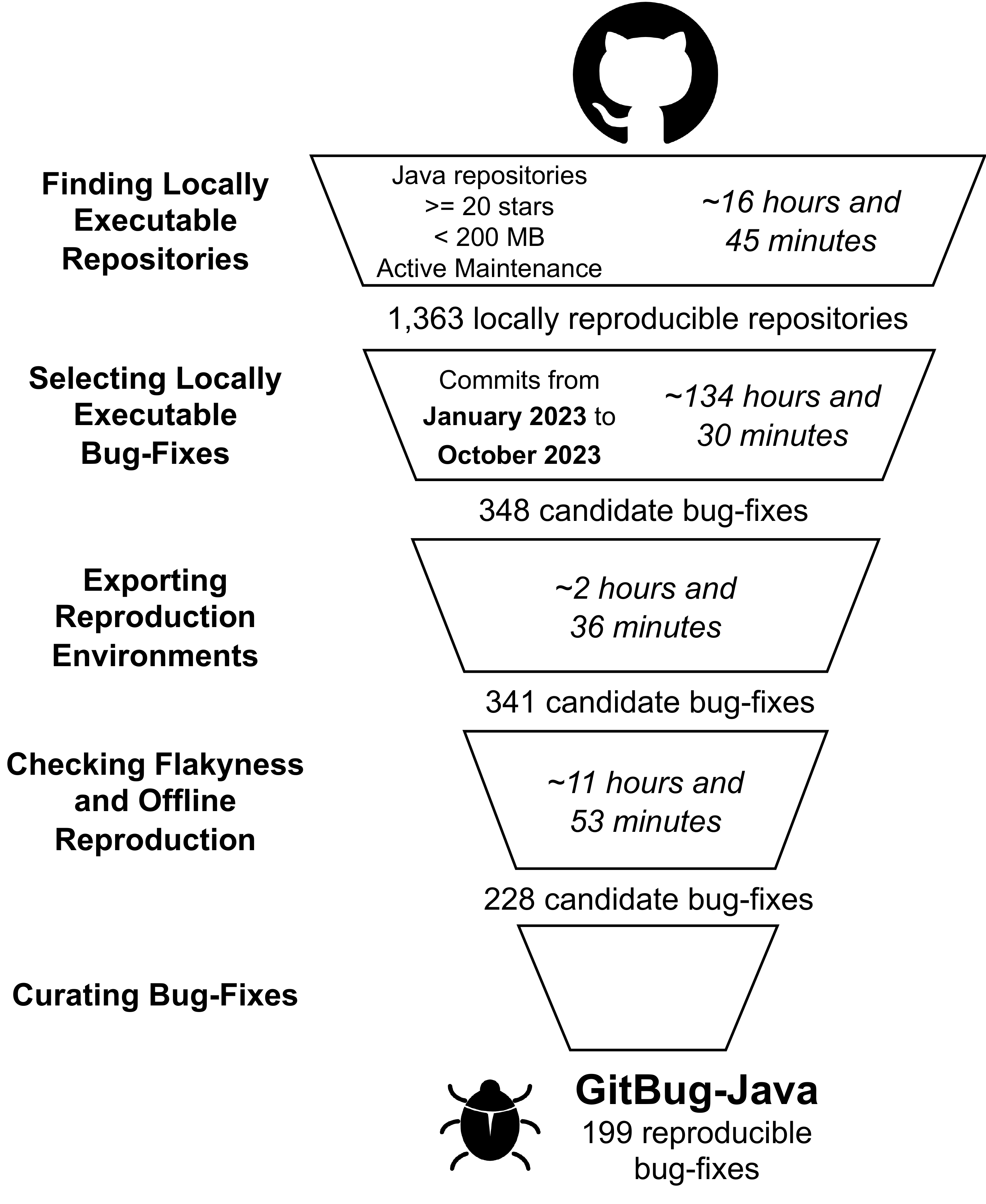}
    \caption{Overview of the building process of \name.}
    \label{fig:overview}
\end{figure}

This section describes the methodology we use to build \name.
The objective is to collect a reproducible Java bug-fix benchmark with recent bugs.
In this paper, Bug-fixes are pairs of subsequent program versions (${program}_{t-1}$, ${program}_{t}$) and a specification given by a test suite $t$, such that $t({program}_{t-1})$ has at least one failing test and $t({program}_{t})$ has no failing test.

Collecting reproducible bug-fixes is a hard task~\cite{zhu2023reproducibility}.
To this end, we employ GitBug-Actions~\cite{saavedra2023gitbug}, a tool for collecting reproducible bug-fix benchmarks using GitHub Actions.
We configure it to look for reproducible bug-fixes in Java projects using Maven or Gradle.
The experimental work is done with 32 parallel workers on a machine with an AMD EPYC 7742 64-Core Processor and 512GB of memory.
The building process starts on November 18th, 2023.
Figure~\ref{fig:overview} shows an overview of the building process of \name.
The full collection artifacts are made available on Zenodo\footnote{\url{https://zenodo.org/records/10579818}}.
We now detail each step of building a reproducible benchmark of recent Java bugs.

\subsection{Finding Locally Executable Repositories}
The first step is to find locally executable repositories.
The objective is to reduce the number of repositories whose commit histories will be searched for reproducible bug-fixes by finding those that meet our quality criteria, including executability.
Recall that, to be locally executable, a repository must, primarily, have GitHub Action workflows that execute tests.

For this, we identify locally executable open-source repositories on GitHub. This allows us to select repositories that meet specific criteria per a GitHub search query\footnote{\url{https://docs.github.com/en/search-github/searching-on-github/searching-for-repositories}}.
We specify the following criteria:
\begin{inparaenum}
    \item Programming Language: Main programming language must be Java per the GitHub metadata.
    \item Popularity: At least 20 stars. The star count serves as an indicator of popularity and community engagement, ensuring that selected repositories to filter out repositories with low relevance and activity.
    \item Size: Repository is less than 200MB. This criterion serves for storage space efficiency.
    \item Active Maintenance: Repositories that are archived are excluded. By excluding archived repositories, we ensure that the selected repositories are relevant and contain recent commits.
\end{inparaenum}

This step takes approximately 16 hours and 45 minutes to complete.
In total, from a universe of 66042 repositories that meet the query criteria, we successfully locally execute 1,363 repositories.

\subsection{Selecting Locally Executable Bug-Fixes}
Next, we look for locally executable bug-fixes in the commit history of each of the locally executable repositories.
We select candidate bug-fixes with two patterns based on their characteristics and test execution results, following \citeauthor{madeiral2019bears}~\cite{madeiral2019bears}:

\begin{inparaenum}
    \item Passing $commit_{t-1}$ + Passing $commit_{t}$ with Source Changes and Test Changes. In this case a developer introduces bug-fix changes as well as test changes to validate them. The test changes are applied to $commit_{t-1}$ and a test failure is expected. Note that, in the case new functionality is introduced, the new tests applied to $commit_{t-1}$ would likely result in an error.

    \item Failing $commit_{t-1}$ + Passing $commit_{t}$ with Only Source Changes. In this case a developer only introduces bug-fix changes such that the program adheres to the already defined specification given by the test suite.
\end{inparaenum}

We only consider commits from January 2023 to October 2023, inclusively.
This step takes approximately 134 hours and 30 minutes to analyze a total of 51,276 commits and results in 508 candidate bug-fixes.
Among these, 160 bug-fixes exclusively modified non-source files and are consequently excluded from the dataset.
In the end, we curate 348 candidate bug-fixes from the initial pool of 51,276 commits.

\subsection{Exporting Reproduction Environments}
Ensuring eternal reproducibility of bug-fixes is one of the main challenges for bug-fix benchmarks~\cite{zhu2023reproducibility}.
The presence of third-party code in applications, obtained during build time, often introduces a dependency on external entities, such as package repositories, whose potential mutability and unavailability pose a threat to the reproducibility of bug benchmarks.
To solve this problem, we build offline reproduction environments to safeguard reproducibility.
The offline reproduction environments are built by saving the Docker's container state after locally executing the tests on the bug-fix commit, preserving all installed software packages required to re-run the tests.
We build the offline reproduction environment for each of the 348 candidate bug-fixes. 
This step takes approximately 2 hours and 36 minutes.
We obtain the offline reproduction environment for 341 out of 348 bug-fixes and are not able to get the reproduction environment for 7 bug-fixes for reasons that include:
\begin{inparaenum}
    \item dependencies that became unavailable between the period when we collected the bug-fixes and when we exported the reproduction environments (this further demonstrates the fragility of reproduction); and
    \item execution time outs, since, for performance reasons, the pipeline uses a default timeout of 10 minutes per test workflow.
\end{inparaenum}
On average, each offline reproduction environment occupies 412.3MiB of storage.

\subsection{Checking Flakiness and Offline Reproduction}
This step verifies that the collected bug-fixes can indeed be reproduced in the reproduction environments previously built.
There are two aspects in verifying that a bug-fix is reproducible.
First, the reproductions must run offline to ensure no third-party is required.
Second, the reproductions must not include flaky tests, which are tests that can intermittently pass or fail~\cite{luo2014empirical}.
This property is undesirable for a bug-fix benchmark like \name.
For instance, an automatic repair tool could apply the correct fix to a bug, but if a flaky test happens to fail while checking if all the tests pass, the tool would receive the information that the fix failed.
To this end, we filter out bug-fixes that are not reproducible.
To do this, we use the reproduction environments to execute each collected bug-fix $5$ times without internet access.
The bug-fixes that yield the exact same test results across all $5$ executions are kept in the benchmark.
This step takes approximately 11 hours and 53 minutes to complete.
There are 100 bug-fixes that either cannot be reproduced offline or are affected by flaky tests in their test suite.
Besides these, a failure also happens to run 13 bug-fixes due to failed Docker operations.
In total, we obtain 228 reproducible bug-fixes. 

\subsection{Curating Bug-Fixes}

We manually curate the candidate bug-fixes, following related work~\cite{madeiral2019bears,just2014defects4j}.
The manual curation serves to validate candidate bug-fixes and remove irrelevant instances.
For example, a candidate bug-fix containing only refactorings and no change to the program's logic is not considered as a relevant bug-fix.
Another example relates to commits that only introduce new features.
The methodology of our manual curation is as follows.
The first and second authors serve as raters.
Each rater rates all candidate bug-fixes, inspecting the commit message, the commit patches, and the failing test information.
For each bug-fix, the rater decides whether the bug-fix is relevant or not to the benchmark.
For the excluded bug-fixes, the raters provide a reason for the exclusion.
After the review process is complete, the raters compare their exclusion lists.
For each bug-fix, if the raters agree on the exclusion, the bug-fix is removed from the benchmark.
If the raters disagree, they discuss whether the bug-fix should be included or not.
If no conclusion is reached, the third author serves as arbiter and decides the inclusion or exclusion of the bug-fix.

This process excludes 29 candidate bug-fixes, resulting in the final set of 199 bug-fixes comprising \name.
15 candidates are removed due to being exclusively refactors.
7 candidates are removed due to exclusively including new features.
6 are removed for being considered irrelevant.
This label is applied to, for example, candidates whose only change is in the string message of an exception.
Lastly, we remove one candidate due to the source code being written in a language other than English.

\section{Insights into \name}

\name contains 199 reproducible Java bug-fixes collected from 55 repositories.
An interactive visualization of \name is published at \url{https://nfsaavedra.github.io/gitbug-java}. 
The full benchmark and command line interface is available on GitHub at \url{https://github.com/gitbugactions/gitbug-java}.

\subsection{Repositories}

\begin{table}
\centering
\caption{Descriptive statistics of the five repositories with the most bugs in \name, and total median/mean across all repositories. \name contains bug-fixes from notable and diverse open-source repositories.}
\label{tab:projects-overview}
\resizebox{\linewidth}{!}{%
\begin{tabular}{cc|crrr|rr} 
\toprule
\multicolumn{2}{c|}{\textbf{Repository}} & \multicolumn{4}{c|}{\textbf{GitHub}\tablefootnote{Information collected in November 22, 2023 during bug-fix collection.}}        & \multicolumn{2}{c}{\textbf{\name}}                                                                                                                 \\
Name                        & Type       & Creation & \multicolumn{1}{c}{Stars} & \multicolumn{1}{c}{\#Commits} & \multicolumn{1}{c|}{\begin{tabular}[c]{@{}c@{}}\#Commits \\2023\end{tabular}} & \multicolumn{1}{c}{\#Tests\tablefootnote{The number of tests per repository is a mean over the bugs included in \name.}} & \multicolumn{1}{c}{\#Bugs}  \\ 
\midrule
traccar/traccar             & App        & 04/2012  & $4431$                    & $8292$                        & $517$                                                                         & $496$                                                                                                                       & $81$                        \\
jhy/jsoup                   & Lib        & 12/2009  & $10392$                   & $1921$                        & $162$                                                                         & $1151$                                                                                                                      & $29$                        \\
jitterted/ensembler         & App        & 04/2021  & $64$                      & $445$                         & $54$                                                                          & $164$                                                                                                                       & $6$                         \\
iipc/jwarc                  & Lib        & 09/2015  & $683$                     & $396$                         & $67$                                                                          & $89$                                                                                                                        & $4$                         \\
snowflakedb/snowflake-jdbc & Lib        & 06/2016  & $163$                     & $1563$                        & $150$                  & $430$                                                                                                                      & $4$                         \\ 
\midrule
Total (Median)              & -          & -        & $106$                     & $395$                         & $52$                                                                          & $221$                                                                                                                       & $1$                         \\
Total (Mean)                & -          & -        & $1569.09$                    & $889.31$                         & $77.71$                                                                          & $511.35$                                                                                                                    & $3.62$                      \\
\bottomrule
\end{tabular}
}
\end{table}

\name contains 199 bug-fixes collected from 55 open-source repositories.
\autoref{tab:projects-overview} shows descriptive statistics of the five projects with the most bugs.
The table reads as follows.
In the first meta-column, we present the name and type of repository where \textit{App} stands for application and \textit{Lib} for library.
In the second meta-column, we present repository information extracted from GitHub including the creation date, the number of stars, the number of commits, and the number of commits from 2023.
In the last meta-column, we present information from \name, namely, the number of tests averaged across all bugs and the number of bugs.
The last two rows present the cumulative statistics of all 55 projects.

These results show that \name includes a diverse range of relevant open-source repositories.
\name includes bug-fixes from relevant repositories created over the span of a decade.
For example, \textit{jhy/jsoup} has been starred by over ten thousand GitHub users.
The commit histories also show that \name includes bug-fixes from repositories with consistent and active development.
For example, \textit{traccar/traccar} has 8292 commits in its history, with 517 of them being from 2023.
In total, the median number of commits per project in the benchmark is 395 (52 from 2023) with a mean of 889 (78 from 2023).
Moreover, the table also suggests that these projects have high-quality test suites.
The median number of tests per project is 221 while the mean is over 511.
Overall, these results show that \name is a benchmark with bug-fixes from diverse but relevant and high-quality sources.

\subsection{Bug-Fixes}

\begin{table}
\centering
\footnotesize
\caption{Descriptive statistics of the size of the bug-fix patch in the number of changed files, hunks, and lines.}
\label{tab:bug-fixes}
\resizebox{\linewidth}{!}{%
\begin{tblr}{
  width = \linewidth,
  colspec = {Q[188]Q[142]Q[183]Q[104]Q[113]Q[167]},
  cells = {r},
  row{1} = {c},
  cell{2}{1} = {c},
  cell{3}{1} = {c},
  cell{4}{1} = {c},
  vline{2} = {-}{},
  hline{1,5} = {-}{0.08em},
  hline{2} = {-}{},
}
        & Mean  & Median & Min & Max & Stddev \\
\#Files & 1.41  & 1      & 1   & 17  & 1.60   \\
\#Hunks & 2.81  & 2      & 1   & 29  & 3.34   \\
\#Lines & 24.63 & 11     & 1   & 576 & 53.22
\end{tblr}
}\end{table}

\name contains 199 reproducible Java bug-fixes.
In this section, we explore key aspects of the bug-fixes.

We compute the number of changed files, unified diff hunks, and lines for each bug-fix patch included in \name.
The number of lines changed is the sum of the added and removed lines.
These provide an insight into the complexity of the changes required to fix \name bugs.
\autoref{tab:bug-fixes} shows descriptive statistics of the number of these changes.
We see that the median patch contains changes in a single file, two hunks, and 9 lines.
The number of changed files ranges from 1 to 17, while hunks range up to 29 and lines up to 586.

\begin{table}
\footnotesize
\centering
\caption{Five most common test failures encountered in the buggy version of \name bug-fixes. The most common failure type is used by libraries such as JUnit5. Long-term reproducibility ensures obtaining the exact same failure over time.}
\label{tab:test-failures}
\resizebox{\linewidth}{!}{%
\begin{tblr}{
  width = \linewidth,
  colspec = {Q[692]Q[223]},
  cells = {c},
  cell{2}{2} = {r},
  cell{3}{2} = {r},
  cell{4}{2} = {r},
  cell{5}{2} = {r},
  cell{6}{2} = {r},
  vline{2} = {-}{},
  hline{1,7} = {-}{0.08em},
  hline{2} = {-}{},
}
Failure Type                        & Proportion \\
org.opentest4j.AssertionFailedError & 52.88\%    \\
java.lang.AssertionError            & 19.81\%    \\
java.lang.NullPointerException      & 7.33\%     \\
org.junit.ComparisonFailure         & 5.00\%     \\
java.lang.IllegalArgumentException  & 4.50\%     
\end{tblr}
}
\end{table}

Another important aspect of bug-fixes are the test suites, and test failures that indicate the presence of a bug and a fix.
The median number of tests in each bug-fix is 497 ($\mu=635.37$, $\sigma=660.01$).
The median number of failing tests in each buggy version is 1 ($\mu=2.38$, $\sigma=7.05$).
We also compute the proportion of test failure types across the whole benchmark.
\autoref{tab:test-failures} shows the five most common test failures encountered in \name bug-fixes.
We see that test assertion errors are the most common (e.g., \textit{AssertionFailedError}), accounting for over 50\% of the failures.

Lastly, we analyze the distribution of bug-fixes over the period considered for bug-fix collection (Jan to Oct 2023).
The median number of bug-fixes per month is 18 ($\mu=20.0$, $\sigma=6.33$).
The minimum number of bug-fixes per month is 14, and the maximum 32.
These results validate \name's methodology, showing that bug-fixes are found across the entire collection period.
Extrapolating from these numbers, future versions of \name could include 240 bugs per year.

\subsection{Reproduction Environments}

\name is made available on GitHub at \url{https://github.com/gitbugactions/gitbug-java}.
The repository includes the scripts and meta-data required to run all 199 bugs included in \name.
A command line interface, alongside documentation, is provided in the repo.
The reproduction environments for offline execution, comprising the files required to rebuild a Docker image for the execution of each bug, are published on Zenodo, with long-term availability provided by CERN, and available through the setup command of the command line interface.

\section{Related Work}

Several bug-fix benchmarks have been proposed by the software engineering community across languages (e.g., FixJS~\cite{csuvik2022fixjs} and BugsJS~\cite{gyimesi2019bugsjs} in Javascript or Defexts in Kotlin \cite{benton2019defexts}).
In this paper, we focus on Java bug-fix benchmarks.
QuixBugs~\cite{lin2017quixbugs}, 
Code4Bench~\cite{majd2019code4bench}, 
RunBugRun~\cite{prenner2023runbugrun}, 
EvalGPTFix~\cite{zhang2023critical} and DebugBench~\cite{tian2024debugbench} are benchmarks constructed from coding competition websites.
Such problems, while real, are not representative of those that developers face in complex software systems.
Others, like HumanEval-Java~\cite{jiang2023impact}, rely on artificially injected bugs which are, by nature, not real.
Benchmarks such as Minecraft~\cite{avulaminecraft} do not include test suites for each bug-fix instance, rendering them unsuitable for studies reliant on execution.
Differently, \name includes executable bug-fixes from real-world software projects.

Defects4J~\cite{just2014defects4j}, Bugs.jar~\cite{saha2018bugs}, Bears~\cite{madeiral2019bears} and BugSwarm~\cite{tomassi2019bugswarm} contain executable bugs from real-world repositories.
Jacontebe \cite{lin2015jacontebe} focuses on concurrency bugs.
We note that Bugs.jar and Bears face serious execution and reproducibility challenges due to missing dependencies and incomplete configuration environments~\cite{zhu2023reproducibility}.
Also, Durieux and Abreu state that 96.4\% of the BugSwarm benchmark is not suitable for APR and FL, for reasons that include duplicate samples, lack of failing tests, and changes to non-source code files~\cite{durieux2019critical}.
Differently, \name only includes fully-reproducible test-suite based bug-fixes.

Finally, Defects4J is a milestone of benchmark research that contains reproducible and adequate bugs for APR and FL.
However, Defects4J mostly contains old bugs.
At the time of writing, it was last updated in 2020.
Indeed, Silva et al.~\cite{silva2021flacoco} find that the majority of Defects4J bugs require Java 6 or earlier bytecode while Java 6 is no longer supported by Oracle as of December 2018.
Moreover, given that the cutoff date of most training datasets used for LLMs is beyond 2020, there exists significant evidence that Defects4J examples are included in them, thus threatening the validity of recent evaluations of LLM-based repair techniques on Defects4J.

GHRB~\cite{lee2023github} is a benchmark that includes 76 real-world Java bug-fixes from after September 2021.
It is designed to contain bug-fixes from after the current OpenAI's cut-off date.
\name is different in two ways.
First, \name is built to be reproducible, shipping the necessary files to build all bugs in offline isolation.
Second, \name is larger and more diverse, containing 199 (+123) bugs from 55 (+39) repositories.

\section{Conclusion}

In this paper, we introduced \name, a reproducible benchmark of recent Java bugs featuring 199 bug-fixes sourced from 55 relevant open-source repositories. 
To ensure the relevance of the bug-fixes to current development practices, we only collected bug-fixes from 2023. \emph{As a result, the freshness of \name bug-fixes mitigates the risk of data and benchmark leakage, a serious concern when evaluating techniques based on large language models (LLMs).}

\name  also provides offline reproduction environments for each collected bug-fix, making \name reproducible in the very long term, even if the dependencies associated with its bug-fixes become unavailable on the Internet. 
To guarantee the validity and quality of the bug-fixes included in \name, we manually curated the included bug-fixes.
\name is publicly available at \url{https://github.com/gitbugactions/gitbug-java}.

\section*{Acknowledgements}
This work was partially supported by the Wallenberg AI, Autono\-mous Systems and Software Program (WASP) funded by the Knut and Alice Wallenberg Foundation.
This work was supported by national funds through FCT, Fundação para a Ciência e a Tecnologia, under grant BD/04736/2023.
This work was supported by national funds through FCT, Fundação para a Ciência e a Tecnologia, under project UIDB/50021/2020 (DOI:10.54499/UIDB/50021/2020).

\bibliographystyle{ACM-Reference-Format}
\balance
\bibliography{main}


\begin{thebibliography}{24}


\ifx \showCODEN    \undefined \def \showCODEN     #1{\unskip}     \fi
\ifx \showDOI      \undefined \def \showDOI       #1{#1}\fi
\ifx \showISBNx    \undefined \def \showISBNx     #1{\unskip}     \fi
\ifx \showISBNxiii \undefined \def \showISBNxiii  #1{\unskip}     \fi
\ifx \showISSN     \undefined \def \showISSN      #1{\unskip}     \fi
\ifx \showLCCN     \undefined \def \showLCCN      #1{\unskip}     \fi
\ifx \shownote     \undefined \def \shownote      #1{#1}          \fi
\ifx \showarticletitle \undefined \def \showarticletitle #1{#1}   \fi
\ifx \showURL      \undefined \def \showURL       {\relax}        \fi
\providecommand\bibfield[2]{#2}
\providecommand\bibinfo[2]{#2}
\providecommand\natexlab[1]{#1}
\providecommand\showeprint[2][]{arXiv:#2}

\bibitem[Avula et~al\mbox{.}(2023)]%
        {avulaminecraft}
\bibfield{author}{\bibinfo{person}{Sai~Krishna Avula}, \bibinfo{person}{Venkatesh Vobbilisetti}, {and} \bibinfo{person}{Shouvick Mondal}.} \bibinfo{year}{2023}\natexlab{}.
\newblock \showarticletitle{Minecraft: Automated Mining of Software Bug Fixes with Precise Code Context}. In \bibinfo{booktitle}{\emph{Proceedings of the 38th IEEE/ACM International Conference on Automated Software Engineering}}.
\newblock


\bibitem[Benton et~al\mbox{.}(2019)]%
        {benton2019defexts}
\bibfield{author}{\bibinfo{person}{Samuel Benton}, \bibinfo{person}{Ali Ghanbari}, {and} \bibinfo{person}{Lingming Zhang}.} \bibinfo{year}{2019}\natexlab{}.
\newblock \showarticletitle{Defexts: A curated dataset of reproducible real-world bugs for modern jvm languages}. In \bibinfo{booktitle}{\emph{2019 IEEE/ACM 41st International Conference on Software Engineering: Companion Proceedings (ICSE-Companion)}}. IEEE, \bibinfo{pages}{47--50}.
\newblock


\bibitem[Csuvik and Vid{\'a}cs(2022)]%
        {csuvik2022fixjs}
\bibfield{author}{\bibinfo{person}{Viktor Csuvik} {and} \bibinfo{person}{L{\'a}szl{\'o} Vid{\'a}cs}.} \bibinfo{year}{2022}\natexlab{}.
\newblock \showarticletitle{FixJS: a dataset of bug-fixing JavaScript commits}. In \bibinfo{booktitle}{\emph{Proceedings of the 19th International Conference on Mining Software Repositories}}. \bibinfo{pages}{712--716}.
\newblock


\bibitem[Durieux and Abreu(2019)]%
        {durieux2019critical}
\bibfield{author}{\bibinfo{person}{Thomas Durieux} {and} \bibinfo{person}{Rui Abreu}.} \bibinfo{year}{2019}\natexlab{}.
\newblock \showarticletitle{Critical review of bugswarm for fault localization and program repair}.
\newblock \bibinfo{journal}{\emph{arXiv preprint arXiv:1905.09375}} (\bibinfo{year}{2019}).
\newblock


\bibitem[Gyimesi et~al\mbox{.}(2019)]%
        {gyimesi2019bugsjs}
\bibfield{author}{\bibinfo{person}{P{\'e}ter Gyimesi}, \bibinfo{person}{B{\'e}la Vancsics}, \bibinfo{person}{Andrea Stocco}, \bibinfo{person}{Davood Mazinanian}, \bibinfo{person}{Arp{\'a}d Besz{\'e}des}, \bibinfo{person}{Rudolf Ferenc}, {and} \bibinfo{person}{Ali Mesbah}.} \bibinfo{year}{2019}\natexlab{}.
\newblock \showarticletitle{Bugsjs: a benchmark of javascript bugs}. In \bibinfo{booktitle}{\emph{2019 12th IEEE Conference on Software Testing, Validation and Verification (ICST)}}. IEEE, \bibinfo{pages}{90--101}.
\newblock


\bibitem[Jacovi et~al\mbox{.}(2023)]%
        {jacovi2023stop}
\bibfield{author}{\bibinfo{person}{Alon Jacovi}, \bibinfo{person}{Avi Caciularu}, \bibinfo{person}{Omer Goldman}, {and} \bibinfo{person}{Yoav Goldberg}.} \bibinfo{year}{2023}\natexlab{}.
\newblock \showarticletitle{Stop uploading test data in plain text: Practical strategies for mitigating data contamination by evaluation benchmarks}.
\newblock \bibinfo{journal}{\emph{arXiv preprint arXiv:2305.10160}} (\bibinfo{year}{2023}).
\newblock


\bibitem[Jiang et~al\mbox{.}(2023)]%
        {jiang2023impact}
\bibfield{author}{\bibinfo{person}{Nan Jiang}, \bibinfo{person}{Kevin Liu}, \bibinfo{person}{Thibaud Lutellier}, {and} \bibinfo{person}{Lin Tan}.} \bibinfo{year}{2023}\natexlab{}.
\newblock \showarticletitle{Impact of Code Language Models on Automated Program Repair}. In \bibinfo{booktitle}{\emph{Proceedings of the 45th International Conference on Software Engineering}} (Melbourne, Victoria, Australia) \emph{(\bibinfo{series}{ICSE '23})}. \bibinfo{publisher}{IEEE Press}, \bibinfo{pages}{1430–1442}.
\newblock
\showISBNx{9781665457019}
\urldef\tempurl%
\url{https://doi.org/10.1109/ICSE48619.2023.00125}
\showDOI{\tempurl}


\bibitem[Just et~al\mbox{.}(2014)]%
        {just2014defects4j}
\bibfield{author}{\bibinfo{person}{Ren{\'e} Just}, \bibinfo{person}{Darioush Jalali}, {and} \bibinfo{person}{Michael~D Ernst}.} \bibinfo{year}{2014}\natexlab{}.
\newblock \showarticletitle{Defects4J: A database of existing faults to enable controlled testing studies for Java programs}. In \bibinfo{booktitle}{\emph{Proceedings of the 2014 international symposium on software testing and analysis}}. \bibinfo{pages}{437--440}.
\newblock


\bibitem[Lee et~al\mbox{.}(2023)]%
        {lee2023github}
\bibfield{author}{\bibinfo{person}{Jae~Yong Lee}, \bibinfo{person}{Sungmin Kang}, \bibinfo{person}{Juyeon Yoon}, {and} \bibinfo{person}{Shin Yoo}.} \bibinfo{year}{2023}\natexlab{}.
\newblock \showarticletitle{The GitHub Recent Bugs Dataset for Evaluating LLM-based Debugging Applications}.
\newblock \bibinfo{journal}{\emph{arXiv preprint arXiv:2310.13229}} (\bibinfo{year}{2023}).
\newblock


\bibitem[Lin et~al\mbox{.}(2017)]%
        {lin2017quixbugs}
\bibfield{author}{\bibinfo{person}{Derrick Lin}, \bibinfo{person}{James Koppel}, \bibinfo{person}{Angela Chen}, {and} \bibinfo{person}{Armando Solar-Lezama}.} \bibinfo{year}{2017}\natexlab{}.
\newblock \showarticletitle{QuixBugs: A multi-lingual program repair benchmark set based on the Quixey Challenge}. In \bibinfo{booktitle}{\emph{Proceedings Companion of the 2017 ACM SIGPLAN international conference on systems, programming, languages, and applications: software for humanity}}. \bibinfo{pages}{55--56}.
\newblock


\bibitem[Lin et~al\mbox{.}(2015)]%
        {lin2015jacontebe}
\bibfield{author}{\bibinfo{person}{Ziyi Lin}, \bibinfo{person}{Darko Marinov}, \bibinfo{person}{Hao Zhong}, \bibinfo{person}{Yuting Chen}, {and} \bibinfo{person}{Jianjun Zhao}.} \bibinfo{year}{2015}\natexlab{}.
\newblock \showarticletitle{Jacontebe: A benchmark suite of real-world java concurrency bugs (T)}. In \bibinfo{booktitle}{\emph{2015 30th IEEE/ACM International Conference on Automated Software Engineering (ASE)}}. IEEE, \bibinfo{pages}{178--189}.
\newblock


\bibitem[Luo et~al\mbox{.}(2014)]%
        {luo2014empirical}
\bibfield{author}{\bibinfo{person}{Qingzhou Luo}, \bibinfo{person}{Farah Hariri}, \bibinfo{person}{Lamyaa Eloussi}, {and} \bibinfo{person}{Darko Marinov}.} \bibinfo{year}{2014}\natexlab{}.
\newblock \showarticletitle{An empirical analysis of flaky tests}. In \bibinfo{booktitle}{\emph{Proceedings of the 22nd ACM SIGSOFT international symposium on foundations of software engineering}}. \bibinfo{pages}{643--653}.
\newblock


\bibitem[Madeiral et~al\mbox{.}(2019)]%
        {madeiral2019bears}
\bibfield{author}{\bibinfo{person}{Fernanda Madeiral}, \bibinfo{person}{Simon Urli}, \bibinfo{person}{Marcelo Maia}, {and} \bibinfo{person}{Martin Monperrus}.} \bibinfo{year}{2019}\natexlab{}.
\newblock \showarticletitle{Bears: An extensible java bug benchmark for automatic program repair studies}. In \bibinfo{booktitle}{\emph{2019 IEEE 26th International Conference on Software Analysis, Evolution and Reengineering (SANER)}}. IEEE, \bibinfo{pages}{468--478}.
\newblock


\bibitem[Majd et~al\mbox{.}(2019)]%
        {majd2019code4bench}
\bibfield{author}{\bibinfo{person}{Amirabbas Majd}, \bibinfo{person}{Mojtaba Vahidi-Asl}, \bibinfo{person}{Alireza Khalilian}, \bibinfo{person}{Ahmad Baraani-Dastjerdi}, {and} \bibinfo{person}{Bahman Zamani}.} \bibinfo{year}{2019}\natexlab{}.
\newblock \showarticletitle{Code4Bench: A multidimensional benchmark of Codeforces data for different program analysis techniques}.
\newblock \bibinfo{journal}{\emph{Journal of Computer Languages}}  \bibinfo{volume}{53} (\bibinfo{year}{2019}), \bibinfo{pages}{38--52}.
\newblock


\bibitem[Prenner and Robbes(2023)]%
        {prenner2023runbugrun}
\bibfield{author}{\bibinfo{person}{Julian~Aron Prenner} {and} \bibinfo{person}{Romain Robbes}.} \bibinfo{year}{2023}\natexlab{}.
\newblock \showarticletitle{RunBugRun--An Executable Dataset for Automated Program Repair}.
\newblock \bibinfo{journal}{\emph{arXiv preprint arXiv:2304.01102}} (\bibinfo{year}{2023}).
\newblock


\bibitem[Saavedra et~al\mbox{.}(2024)]%
        {saavedra2023gitbug}
\bibfield{author}{\bibinfo{person}{Nuno Saavedra}, \bibinfo{person}{Andr{\'e} Silva}, {and} \bibinfo{person}{Martin Monperrus}.} \bibinfo{year}{2024}\natexlab{}.
\newblock \showarticletitle{GitBug-Actions: Building Reproducible Bug-Fix Benchmarks with GitHub Actions}. In \bibinfo{booktitle}{\emph{2024 IEEE/ACM 46th International Conference on Software Engineering: Companion Proceedings (ICSE-Companion)}}. IEEE.
\newblock


\bibitem[Saha et~al\mbox{.}(2018)]%
        {saha2018bugs}
\bibfield{author}{\bibinfo{person}{Ripon~K Saha}, \bibinfo{person}{Yingjun Lyu}, \bibinfo{person}{Wing Lam}, \bibinfo{person}{Hiroaki Yoshida}, {and} \bibinfo{person}{Mukul~R Prasad}.} \bibinfo{year}{2018}\natexlab{}.
\newblock \showarticletitle{Bugs. jar: A large-scale, diverse dataset of real-world java bugs}. In \bibinfo{booktitle}{\emph{Proceedings of the 15th international conference on mining software repositories}}. \bibinfo{pages}{10--13}.
\newblock


\bibitem[Silva et~al\mbox{.}(2021)]%
        {silva2021flacoco}
\bibfield{author}{\bibinfo{person}{Andr{\'e} Silva}, \bibinfo{person}{Matias Martinez}, \bibinfo{person}{Benjamin Danglot}, \bibinfo{person}{Davide Ginelli}, {and} \bibinfo{person}{Martin Monperrus}.} \bibinfo{year}{2021}\natexlab{}.
\newblock \showarticletitle{Flacoco: Fault localization for java based on industry-grade coverage}.
\newblock \bibinfo{journal}{\emph{arXiv preprint arXiv:2111.12513}} (\bibinfo{year}{2021}).
\newblock


\bibitem[Sim et~al\mbox{.}(2003)]%
        {sim2003using}
\bibfield{author}{\bibinfo{person}{Susan~Elliott Sim}, \bibinfo{person}{Steve Easterbrook}, {and} \bibinfo{person}{Richard~C Holt}.} \bibinfo{year}{2003}\natexlab{}.
\newblock \showarticletitle{Using benchmarking to advance research: A challenge to software engineering}. In \bibinfo{booktitle}{\emph{25th International Conference on Software Engineering, 2003. Proceedings.}} IEEE, \bibinfo{pages}{74--83}.
\newblock


\bibitem[Tian et~al\mbox{.}(2024)]%
        {tian2024debugbench}
\bibfield{author}{\bibinfo{person}{Runchu Tian}, \bibinfo{person}{Yining Ye}, \bibinfo{person}{Yujia Qin}, \bibinfo{person}{Xin Cong}, \bibinfo{person}{Yankai Lin}, \bibinfo{person}{Zhiyuan Liu}, {and} \bibinfo{person}{Maosong Sun}.} \bibinfo{year}{2024}\natexlab{}.
\newblock \showarticletitle{DebugBench: Evaluating Debugging Capability of Large Language Models}.
\newblock \bibinfo{journal}{\emph{arXiv preprint arXiv:2401.04621}} (\bibinfo{year}{2024}).
\newblock


\bibitem[Tomassi et~al\mbox{.}(2019)]%
        {tomassi2019bugswarm}
\bibfield{author}{\bibinfo{person}{David~A Tomassi}, \bibinfo{person}{Naji Dmeiri}, \bibinfo{person}{Yichen Wang}, \bibinfo{person}{Antara Bhowmick}, \bibinfo{person}{Yen-Chuan Liu}, \bibinfo{person}{Premkumar~T Devanbu}, \bibinfo{person}{Bogdan Vasilescu}, {and} \bibinfo{person}{Cindy Rubio-Gonz{\'a}lez}.} \bibinfo{year}{2019}\natexlab{}.
\newblock \showarticletitle{Bugswarm: Mining and continuously growing a dataset of reproducible failures and fixes}. In \bibinfo{booktitle}{\emph{2019 IEEE/ACM 41st International Conference on Software Engineering (ICSE)}}. IEEE, \bibinfo{pages}{339--349}.
\newblock


\bibitem[Wright et~al\mbox{.}(2010)]%
        {wright2010validity}
\bibfield{author}{\bibinfo{person}{Hyrum~K Wright}, \bibinfo{person}{Miryung Kim}, {and} \bibinfo{person}{Dewayne~E Perry}.} \bibinfo{year}{2010}\natexlab{}.
\newblock \showarticletitle{Validity concerns in software engineering research}. In \bibinfo{booktitle}{\emph{Proceedings of the FSE/SDP workshop on Future of software engineering research}}. \bibinfo{pages}{411--414}.
\newblock


\bibitem[Zhang et~al\mbox{.}(2023)]%
        {zhang2023critical}
\bibfield{author}{\bibinfo{person}{Quanjun Zhang}, \bibinfo{person}{Tongke Zhang}, \bibinfo{person}{Juan Zhai}, \bibinfo{person}{Chunrong Fang}, \bibinfo{person}{Bowen Yu}, \bibinfo{person}{Weisong Sun}, {and} \bibinfo{person}{Zhenyu Chen}.} \bibinfo{year}{2023}\natexlab{}.
\newblock \bibinfo{title}{A Critical Review of Large Language Model on Software Engineering: An Example from ChatGPT and Automated Program Repair}.
\newblock
\newblock
\showeprint[arxiv]{2310.08879}~[cs.SE]


\bibitem[Zhu and Rubio-Gonz{\'a}lez(2023)]%
        {zhu2023reproducibility}
\bibfield{author}{\bibinfo{person}{Hao-Nan Zhu} {and} \bibinfo{person}{Cindy Rubio-Gonz{\'a}lez}.} \bibinfo{year}{2023}\natexlab{}.
\newblock \showarticletitle{On the reproducibility of software defect datasets}.
\newblock \bibinfo{journal}{\emph{ICSE. IEEE}} (\bibinfo{year}{2023}).
\newblock


\end{thebibliography}

\end{document}